\documentclass[%
reprint,
superscriptaddress,
 prl,
]{revtex4-2}

\usepackage{graphicx}
\usepackage{dcolumn}
\usepackage{bm}
\usepackage{hyperref}
\usepackage[mathlines]{lineno}
\usepackage[T1]{fontenc}
\usepackage{amsmath}
\usepackage{natbib}

\usepackage{xcolor}

\begin{document}

\preprint{APS/123-QED}

\title{Observation of degenerate zero-energy topological states\\ at disclinations in an acoustic lattice}

\author{Yuanchen Deng}
    \affiliation{Graduate Program in Acoustics,  The Pennsylvania State University, University Park, Pennsylvania, 16802, USA}
\thanks{Y.D., W.A.B. and Z.C. contributed equally to this work.}    
\author{Wladimir A. Benalcazar}
    \email{wb7707@princeton.edu}
    \affiliation{Department of Physics, Princeton University, Princeton, NJ, 08542, USA}
    \affiliation{Department of Physics, The Pennsylvania State University, University Park, Pennsylvania 16802, USA} 
\thanks{Y.D., W.A.B. and Z.C. contributed equally to this work.} 

\author{Ze-Guo Chen}
    \affiliation{Department of Physics, Hong Kong Baptist University, Kowloon Tong, Hong Kong, China}
\thanks{Y.D., W.A.B. and Z.C. contributed equally to this work.} 

\author{Mourad Oudich}
    \affiliation{Graduate Program in Acoustics,  The Pennsylvania State University, University Park, Pennsylvania, 16802, USA}
    \affiliation{Université de Lorraine, CNRS, Institut Jean Lamour, F-54000 Nancy, France} 
\author{Guancong Ma}
    \email{phgcma@hkbu.edu.hk}
    \affiliation{Department of Physics, Hong Kong Baptist University, Kowloon Tong, Hong Kong, China}
\author{Yun Jing}
    \email{yqj5201@psu.edu}
    \affiliation{Graduate Program in Acoustics,  The Pennsylvania State University, University Park, Pennsylvania, 16802, USA}
\date{\today}

\begin{abstract}
Building upon the bulk-boundary correspondence in topological phases of matter, disclinations have recently been harnessed to trap fractionally quantized density of states (DoS) in classical wave systems. While these fractional DoS have associated states localized to the disclination's core, such states are not protected from deconfinement due to the breaking of chiral symmetry, generally leading to resonances which, even in principle, have finite lifetimes and suboptimal confinement. Here, we devise and experimentally validate in acoustic lattices a paradigm by which topological states bind to disclinations without a fractional DoS but which preserve chiral symmetry. The preservation of chiral symmetry pins the states at the mid-gap, resulting in their protected maximal confinement. The integer DoS at the defect results in two-fold degenerate states that, due to symmetry constraints, do not gap out. Our study provides a fresh perspective on the interplay between symmetry-protection in topological phases and topological defects, with possible applications in classical and quantum systems alike.
\end{abstract}

\maketitle
Although originally conceived to explain electronic quantum phases of matter~\cite{Klitzing1980,Thouless1982}, topological band theory applies to wave phenomena at large~\cite{Haldane2008}. Hence, it is relevant to a wide range of classical systems and has recently found fertile ground in acoustic~\cite{Zhang2018, Ma2019a}, mechanical~\cite{Ma2019a}, and photonic platforms~\cite{Ozawa2019a}. For example, it provides mechanisms for the generation of robust one-way states~\cite{Hafezi2013a,Rechtsman2013}, topologically robust corner states~\cite{benalcazar2017, Noh2018,Xue2019,Ni2019}, symmetry-protected bound states in the continuum~\cite{Marinica2008,Plotnik2011,Hsu2016}, topological wave steering~\cite{He2018,Xu2021}, and topological lasers~\cite{Harari2018,Bandres2018}.

At the core of these phenomena is the existence of robust in-gap states, which are protected by a \emph{bulk-boundary correspondence}; if the bulk of the material is topological, in-gap states robust against perturbations or deformations will exist at its boundaries, as long as certain symmetries are preserved. An important extension of this principle applies to specific topological defects, where the existence of topological states hinges on an interplay between the bulk topology of the lattice and the topological charge of the defect~\cite{Ran2009,Teo2010,Teo2013,Benalcazar2014,Li2020a}. Notable examples include topological states bound to vortices~\cite{Gao2019,Iadecola2016,Noh2020,Menssen2020}, dislocations~\cite{Hughes2014,Paulose2015,Ye2021,Xue2021a}, and disclinations~\cite{Li2020,Wang2021,Liu2021,Peterson2021,Wang2020,Xie2021}. 

Topological defects allows binding topological states within the bulk --as opposed to the boundaries-- in periodic synthetic platforms. 
These states are particularly beneficial if they lie at mid-gap, as this guarantees both spectral isolation and maximal confinement, which in turn maximizes nonlinear effects and wave-matter interaction for sensing purposes. In order to pin topological states to mid-gap, chiral symmetry must be preserved. Unfortunately, in many classical systems, dislocations and disclinations often disrupt chiral symmetry as they destroy the bipartite nature of chiral-symmetric lattices. Consequently, topological states associated with these defects are not protected from deconfinement. This is the case of the recently realized disclination states of Refs.~\onlinecite{Liu2021,Peterson2021}. In them, a topological fractional density of states protect states bound to disclinations. However, the reported associated states are either (i) hybridized with bulk states forming resonances~\cite{Liu2021} or (ii) bound states fine-tuned to be \emph{in-gap} but not protected by symmetry to be at \emph{mid-gap}~\cite{Peterson2021}.

In this work, we demonstrate in theory and experiments how states bound to the core of disclinations can be symmetry-protected to lie at mid-gap in certain obstructed atomic limit (OAL) topological phases, thus ensuring spectral isolation of these states from bulk states and maximizing their confinement to the defect's core. The underlying protection mechanism arises not from the interplay of bulk and defect topologies~\cite{Liu2021,Peterson2021}, but from the interplay of chiral symmetry in the lattice at large, the point group symmetry of the topological defect, and the topological phase of the lattice. Said succinctly, the protection mechanism rests on the fact that zero energy states with opposite chiral charges --which can generally hybridize into the bulk-- are prevented from doing so when they form a two-dimensional irreducible representation of some point group symmetry. The point group symmetry of the defect forces the states to be degenerate, and chiral symmetry forces them to be pinned at mid-gap. 

To demonstrate this protection mechanism, we have devised acoustic lattices that preserve a homogeneous coupling strength across the lattice despite the curvature induced by the disclinations. We implement our protection mechanism in this acoustic system and present the first experimental observation of degenerate, symmetry-protected, mid-gap states at the core of topological defects in synthetic platforms. Our ability to protect multiple degenerate topological states at a single topological defect further advances the technological relevance of these states, as it increases the density of states available for lasing~\cite{Harari2018,Bandres2018} or coupling to external devices.

\begin{figure}
\includegraphics[width=8cm]{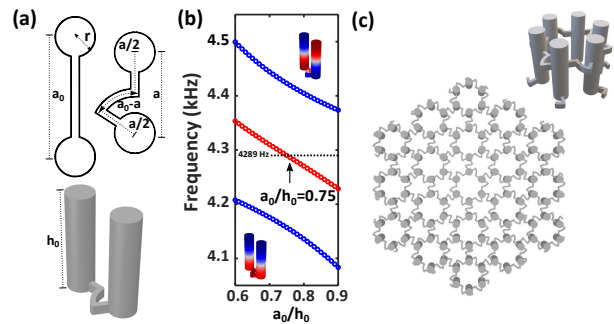}
\caption{\label{fig:1}(a) Top: the top view of a conventional straight tube coupled-cavity model and a coiled tube coupled-cavity model. Bottom: a 3D view of a coiled tube coupled-cavity model. (b) Frequency spectrum of the coupled cavity system with coiled coupling tubes shown in (a), bottom, as a function of the ratio between $a_0/h_0$. The blue markers represent the frequencies of the symmetric modes (lower frequency) and the anti-symmetric modes (higher frequency).The average frequencies of the symmetric and anti-symmetric modes are marked with red circles. The symmetric and anti-symmetric modes are distributed symmetrically about the zero-energy level(4289Hz) at $a_0/h_0$=0.75, indicating chiral symmetry. 
(c) The top-view of the OAL(3c) honeycomb lattice with coiled coupling tubes, where the external coupling tubes are situated at the bottom of the cavity and the internal coupling tubes are located at $h_0$/4 above the center of the cavity. The figure on the top right shows a single unit cell with coiled coupling tubes.}
\end{figure}

\begin{figure}
\includegraphics[width=8cm]{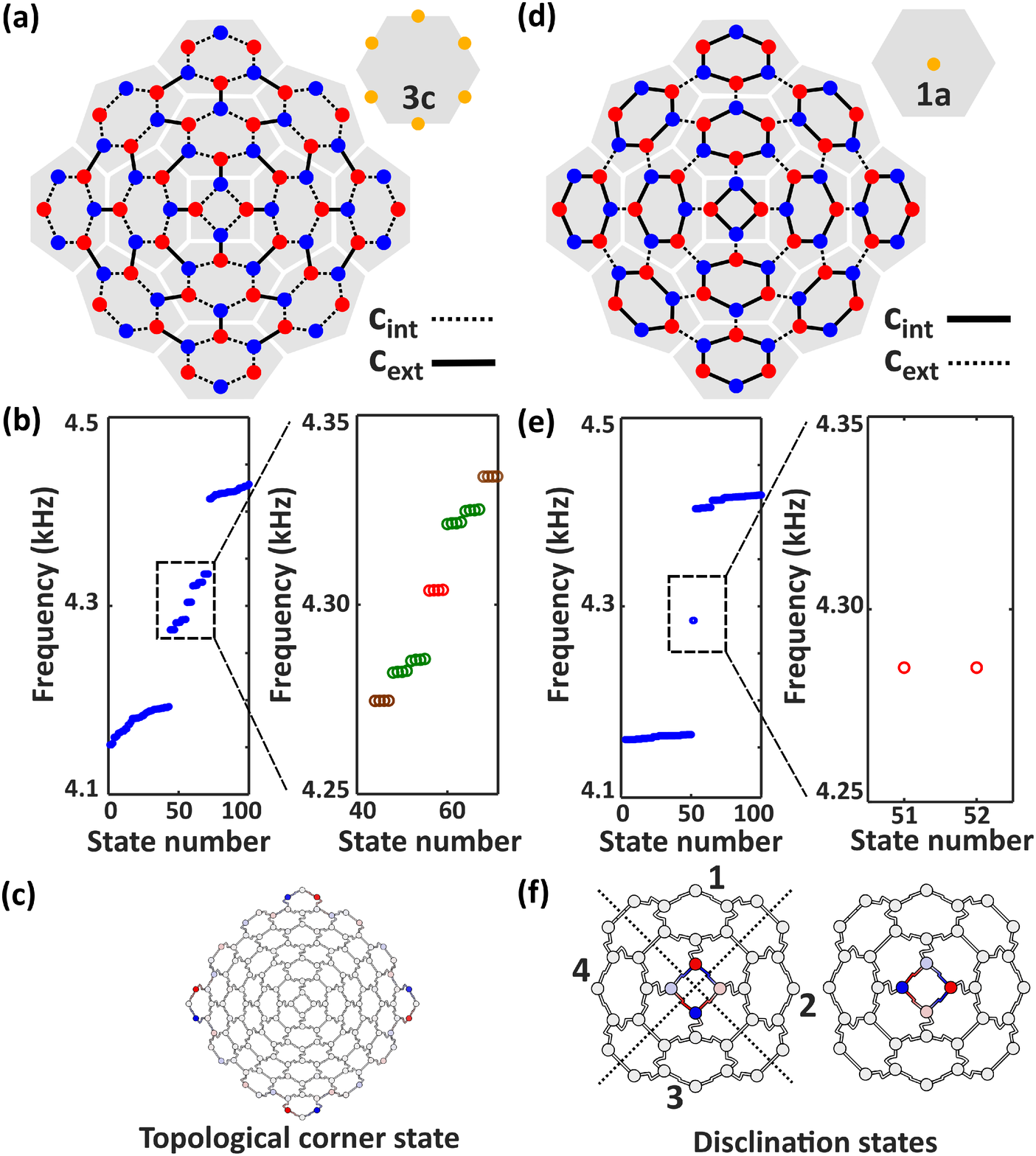}
\caption{\label{fig:2}(a) The OAL(3c) lattice with a $2\pi/3$ disclination. Each shade represents one unit cell. Two different sublattices are distinguished by red and blue circles. The inset figure shows a unit cell with its Wannier centers at half-filling at Wyckoff position 3c. (b) Numerically computed eigenfrequencies for the OAL(3c) structure. The topological corner states, edge states, and trivial corner states are represented by red, green, and brown circles, respectively. (c) The four degenerate topological corner states at 4304 Hz. (d) The OAL(1a) lattice with a $2\pi/3$ disclination. The inset figure shows a unit cell with its Wannier centers at half-filling at Wyckoff position 1a (three-fold degenerate). (e) Numerically computed eigenfrequencies for the OAL(1a) structure. (f) The pair of degenerate disclination bound states at 4285 Hz. The dotted lines highlight the quadrants. Only the region surrounding the lattice core is shown for better visualization.}
\end{figure}

Our acoustic lattice relies on a coupled-cavity acoustic model~\cite{Ding2016,Yang2018,Chen2020}. As shown in Fig.~\ref{fig:1}(a), two identical cylindrical cavities with radius $r = 0.5$ cm are coupled via a tube with a deep sub-wavelength cross-section. The length of the tube plus the diameter of the cavity is $a_0$ and the height of the cavity is $h_0 = 4$ cm. Here, the first-order resonance (4289Hz), which has a cosine-function acoustic profile along the cavity’s axial direction with one nodal plane in the middle, is used as the onsite orbital. To produce a chiral symmetric system, the ratio between $a_0$ and $h_0$ is set at an optimal value $0.75$ based on the eigenmode analysis using COMSOL Multiphysics, as shown in Fig.~\ref{fig:1}(b)~\cite{Chen2020}. An acoustic honeycomb lattice is then constructed as shown in Fig.~\ref{fig:1}(c). Our acoustic model has two salient features that enable us to investigate the symmetry-protected disclination states. First, the coupling tube can be coiled [Fig.~\ref{fig:1}(a)] while preserving its coupling strength. This feature stems from the fact that only the fundamental mode is permitted in these subwavelength channels. It then follows that the coupling is dictated by the total length of the tube rather than the separation between two cavities. Such a coiling mechanism is vital for studying deformed lattices since it allows the arbitrary placement of \emph{atoms} while maintaining a homogeneous coupling strength throughout the entire lattice. As such, this system is well-suited to implementing disclinations, which induce curvature singularities that result in geometric distortions when projected onto flat surfaces. Second, the coupling among cavities is proportional to the local acoustic amplitudes in the cavities, which follows a cosine function along the cavity's axial length. Thus the ratio between couplings within a unit cell ($c_\text{int}$) and couplings among neighboring unit cells ($c_\text{ext}$) is tunable by the position of the external and internal coupling tubes~\cite{SuppMat1}. We construct the honeycomb lattice with Kekule modulations of the couplings to generate two obstructed atomic limit (OAL) topological phases, both of which are chiral symmetric~\cite{VanMiert2018,Noh2018,Proctor2020} [Fig.~\ref{fig:2}(a) and \ref{fig:2}(b)]. The Kekule modulation consists of having two different couplings: $c_\text{int}$ within unit cells, and $c_\text{ext}$ among neighboring unit cells. When $c_\text{int}$<$c_\text{ext}$, the lattice is in an OAL phase with Wannier centers at Wyckoff position $3c$ of the unit cell as shown in the inset figure of Fig.~\ref{fig:2}(a) (See Supplementary Material~\cite{SuppMat1} for more details on the OAL(3c) lattice). On the other hand, when $c_\text{int}$> $c_\text{ext}$, the lattice is in an OAL phase with three Wannier centers at Wyckoff position $1a$ of the unit cell at \emph{half-filling} as shown in the inset figure of Fig.~\ref{fig:2}(d). While in both phases the bulk polarization~\cite{vanderbilt1993} vanishes due to the presence of $C_6$ symmetry~\cite{Fang2012,Benalcazar2019}, the OAL(3c) phase has nontrivial second-order topological index~\cite{VanMiert2018,Benalcazar2019}. At $c_\text{int}$=$c_\text{ext}$, the lattice is in the perfect honeycomb configuration. 

We introduce a disclination to the honeycomb lattices by the Volterra process of removing a $2\pi/3$ section of a hexagonal sample~\cite{Teo2013a, SuppMat1}. Such a process generates a disclination with a Frank angle of $2\pi/3$ and an overall $C_{4v}$ symmetric structure, with the center of rotation at the core of the disclination. The curvature singularity deforms the lattices as shown in Figs.~\ref{fig:2}(a) and \ref{fig:2}(d) for both OAL phases. To counter the effect of this deformation on the couplings, the coupling tubes are coiled at each site to ensure a uniform overall length (and thus coupling strength) across the entire lattice via the mechanism introduced earlier. Our configuration sacrifices a fractional density of states at the disclination in the OAL(3c) phase (which is obtained in the same lattice but with a $\pi/3$ disclination~\cite{Benalcazar2019,Liu2021,Peterson2021} which was the first one to predict this fractionalization of the density of states), in favor of preserving chiral symmetry~\cite{SuppMat1}. As we will show, the presence of chiral symmetry and either $C_4$ and time-reversal or $C_{4v}$ symmetry protect two degenerate states at the core of the disclinations in only one of the two OAL phases.
 
Chiral-symmetric systems have Hamiltonians $h$ that obey $\Pi h\Pi^{-1}=-h$, where $\Pi$ is the chiral operator. For every eigenstate $\psi$ of $h$ with energy $\epsilon$ (such that $ h \psi = \epsilon \psi$), there is a second eigenstate $\Pi \psi$ with energy $-\epsilon$. This can easily be seen by operating $h \Pi \psi = -\Pi h \psi = -\epsilon \Pi \psi$. Thus, the energies in a system with chiral symmetry come in pairs $(\epsilon,-\epsilon)$, and their states are related by the chiral-symmetry operator $\Pi$.
Our acoustic model is composed of a lattice with 4 unit cells per side. The OAL(3c) lattice hosts 4 topological corner states at zero-energy, as shown in Figs.~\ref{fig:2}(b) and~\ref{fig:2}(c) and reported earlier in Ref.~\cite{Noh2018}. The symmetry of the spectrum indicates that the lattice with disclination preserves chiral symmetry. Further confirmation comes from the fact that the corner states have support only on one sublattice at each corner, indicating that they are eigenstates of the chiral operator with well-defined chiral charges and thus are zero-energy states. The OAL(1a) lattice, on the other hand, does not possess corner states. However, it possesses a pair of mid-gap degenerate states confined to the disclination core, as illustrated in Figs.~\ref{fig:2}(e) and~\ref{fig:2}(f). These states are originally presented in this work and are the main finding of our paper. Tight binding model (TBM) simulation results show similar mode distributions in the bandgaps, which corroborates our COMSOL simulation results~\cite{SuppMat1}. These two states have support over both sublattices, indicating that their overall chiral charge is zero; however, these two states form a 2D irreducible representation (irrep) of $C_4$ plus time-reversal symmetry (TRS) or $C_{4v}$ (i.e., $C_4$ symmetry plus reflection symmetry), which prevents them from being lifted away from the zero-energy level. The 2D irrep of $C_{4v}$ or $C_4$ and TRS can be described by the basis ${\psi_+}=\frac{1}{2}(1,i,-1,-i)^T$ and ${\psi_-}=\frac{1}{2}(1,-i,-1,i)^T$, where the entries correspond to the disclination's sites at each quadrant, respectively, as shown in Fig.~\ref{fig:2}(f), top (the states in Fig.~\ref{fig:2}(f) are proportional to $\frac{1}{\sqrt{2}}(\psi_+ \pm \psi_-)$). The states $\psi_\pm$ are eigenstates of $C_4$ and map to one another under TRS or reflection symmetry. Since $\psi_\pm$ are a basis for the 2D irrep, they must remain degenerate in energy as long as the above-mentioned symmetries are preserved. This basis is convenient because, in the presence of chiral symmetry, $\psi_\pm$ are chiral partners of each other, i.e., $\psi_+=\Pi \psi_-$ and vice versa, from which it follows that these two states should have energies of opposite sign, $\epsilon, -\epsilon$. Thus, under $C_{4v}$ symmetry or $C_4$ symmetry plus TRS, as well as chiral symmetry, $\psi_\pm$ must both have $\epsilon=0$ identically. In contrast, the OAL(3c) phase does not enclose the 2D representation at the core (only at its corners), and thus it does not trap zero-energy states at the disclination core. 

\begin{figure}
\includegraphics[width=8cm]{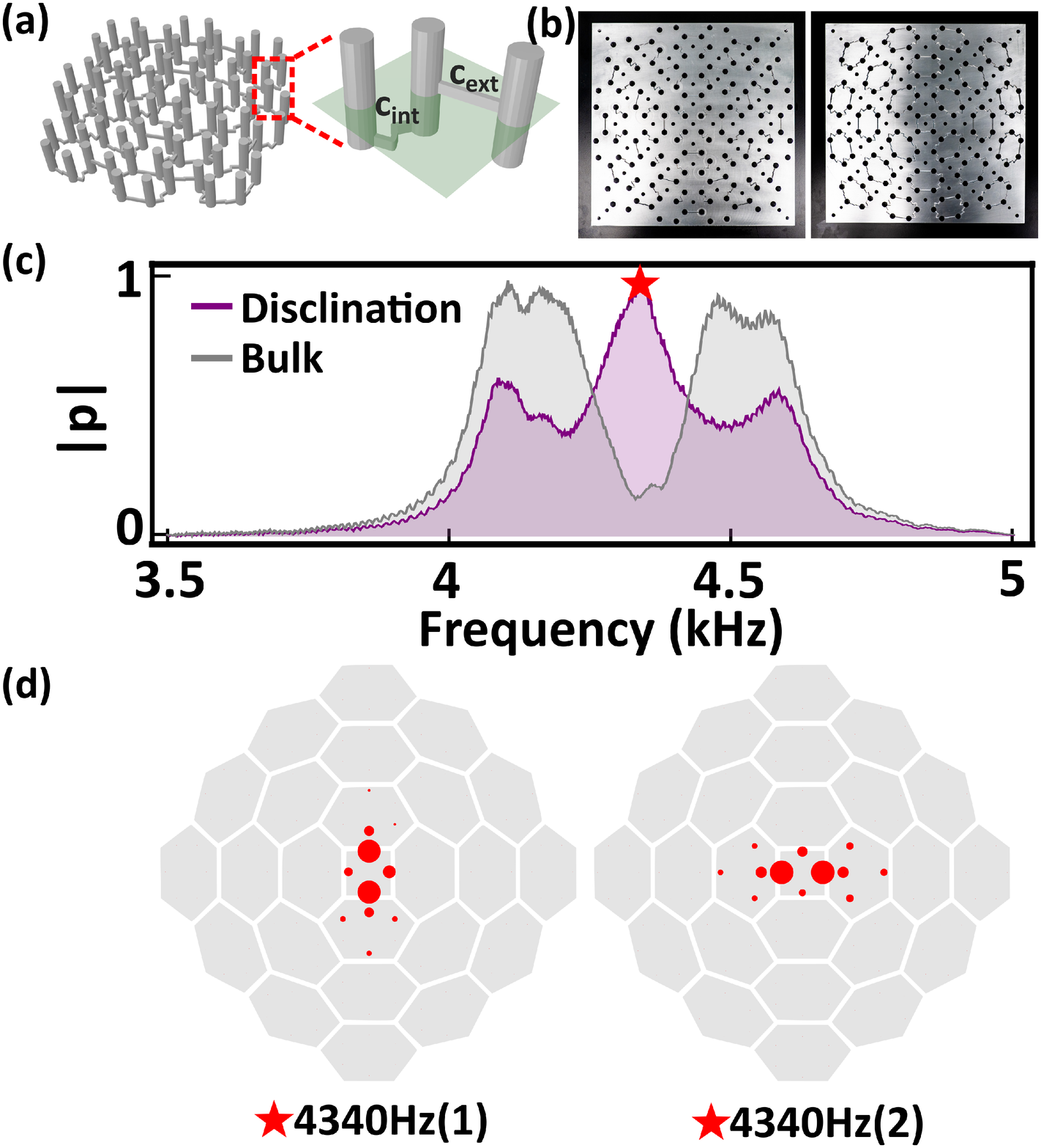}
\caption{\label{fig:3}(a) (top panel) The acoustic OAL(1a) lattice. Only the inner 3 by 3 unit cells are shown here for better visualization. (bottom panel) A close-up view of three cavities in the dashed line box shows the position of the external and internal coupling tubes. The transparent cut-plane indicates the interface between the two layers used to construct the experimental acoustic sample. (b) Photographs of the OAL(3c) acoustic lattice sample with its cavities (the larger holes) and coupling channels. The two blocks are stacked and then sealed to form the coupled-cavity lattice. The smaller holes without tubes are for mounting purposes. (c) Spectra of the normalized pressure amplitude $|p|$ of the disclination (purple) and bulk (grey) states. The degenerate disclination states are marked with the red star (two degenerate states at $4340$ Hz). (d) The pressure distribution maps of the two disclination states at the frequency marked by the red star in (c). The area of the circle represents the amplitude of the pressure. Note that the entire lattice is measured, but the pressure amplitudes are too weak away from the disclination core.}
\end{figure}

We have experimentally measured two samples corresponding to the OAL(1a) and OAL(3c) lattices containing the $2\pi/3$ disclination. Only the results of the OAL(1a) lattice are discussed here, while the OAL(3c) results showing corner states can be found in the Supplementary Material~\cite{SuppMat1}. An illustration of the OAL(1a) acoustic lattice is shown in Fig.~\ref{fig:3}(a). The internal and external coupling tubes are machined on two separate aluminum blocks as shown in Fig.~\ref{fig:3}(b) and then stacked together. We measure both the bulk and disclination responses of the acoustic lattice, and the results are shown in Fig.~\ref{fig:3}(c). Details of the experiment can be found in the Supplementary Material~\cite{SuppMat1}. The bulk spectrum shows a gap around $4.3$ kHz, while the disclination core response shows a single peak located at the mid-gap and two lower peaks within the bulk band frequencies. The symmetry of the two spectra around mid-gap is a signature of the well-preserved chiral symmetry in the acoustic lattice. We then raster-map the response profile in the entire lattice by measuring the pressure amplitude at the top of each cavity. The results show that the mid-gap peak indeed corresponds to the pair of degenerate, symmetry-protected disclination bound states as shown in Fig.~\ref{fig:3}(d). These two states are at 4340 Hz, slightly off from the numerically predicted frequency (4285 Hz) due to fabrication variations. The degenerate disclination bound states are orthogonal to one another, and thus they must be separately excited. The other two lower peaks within the bulk band frequencies are resulted from two states at the disclination which are orthogonal to mid-gap ones, and are not maximally localized.

Since the symmetry representations of the states within a topological phase in the lattice are stable as long as the symmetries are preserved, our protection mechanism is robust to symmetry-preserving perturbations. In a chiral-symmetric lattice, our zero energy states can be removed from the core only upon a topological phase transition from the OAL(1a) phase to the OAL(3c) phase, where a reconfiguration of the irreps occurs (the 2D irrep of the zero states moves from the disclination core to the corner states). To examine the robustness of the disclination bound states, we have conducted additional simulations with different types of perturbations to the disclination core, and the results can be found in the Supplementary Material~\cite{SuppMat1}, along with additional discussion on the protection mechanism of the zero-energy disclination modes. 

In conclusion, we have theoretically and experimentally studied a mechanism that protects the maximal confinement of states at topological defects with chiral symmetry. Our mechanism relies on the interplay of the point group symmetry of the topological defect and the topological phase of a lattice. The sonic mid-gap disclination states could not only inspire new routes for controlling acoustic local density of states for sound emission control~\cite{Landi2018a}, but also paves the way of novel energy transportation mechanisms via topological disclination pumps~\cite{Xie2021}. In addition to acoustics, our theory can be applied to general wave phenomena (electromagnetic and elastic wave systems) and is equally applicable to quantum systems in condensed matter physics.

\begin{acknowledgements}
Y.J. thanks the support from NSF through CMMI-1951221 and CMMI-2039463. W.A.B. thanks the support of the Moore Postdoctoral Fellowship at Princeton University and the Eberly Postdoctoral Fellowship at the Pennsylvania State University. G. M. is supported by Hong Kong Research Grants Council (12302420, 12300419, 22302718, C6013-18G), National Natural Science Foundation of China Excellent Young Scientist Scheme (Hong Kong \& Macau) (11922416) and Youth Program (11802256).
\end{acknowledgements}




%

\end{document}